\documentstyle[12pt,epsfig]{article}

\newcommand{\f}{\begin{equation}}
\newcommand{\ff}{\end{equation}}
\newcommand{\fa}{\begin{eqnarray}}
\newcommand{\ffa}{\end{eqnarray}}

\title{Holographic Formulation of Quantum
Supergravity}
\author{Yi Ling$^{\dag}$ and Lee Smolin$^{*}$
\thanks{email:\dag
ling@phys.psu.edu ,$^{*}$ smolin@phys.psu.edu}\\ \centerline{\it
Center for Gravitational Physics and Geometry}\\ \centerline{\it
Department of Physics}\\ \centerline {\it The Pennsylvania State
University}\\ \centerline{\it University Park, PA, USA 16802}\\
\centerline{and} \\ \centerline{\it The Blackett Laboratory}
\\ \centerline{\it Imperial College of Science, Technology and
Medicine}\\ \centerline{\it South Kensington, London SW7 2BZ, UK}}
\begin{document}
\maketitle
\begin{abstract}
\baselineskip=20pt

We show that ${\cal N}=1$ supergravity with a cosmological
constant  can be expressed as constrained topological field theory
based on the supergroup $Osp(1|4)$. The theory is then extended to
include timelike boundaries with finite spatial area.  Consistent
boundary conditions are found which induce a boundary theory based
on a supersymmetric Chern-Simons theory. The boundary state space
is constructed from states of the boundary supersymmetric
Chern-Simons theory on the punctured two sphere and naturally
satisfies the Bekenstein bound, where area is measured by the area
operator of quantum supergravity.
\end{abstract}
\eject\tableofcontents \eject
\section{Introduction}
\baselineskip=20pt In this paper we describe a formulation of
quantum supergravity with a finite cosmological constant, in the
presence of a timelike boundary with finite spatial area. We show
that, as in the case of quantum general relativity
\cite{linking,hologr}, a holographic formulation of the theory
naturally emerges. This work is part of a series of papers
\cite{superspin,supern=2,holon=2} in which we are extending to
quantum supergravity the methods \cite{sn1,sn2,spain,lp1,lp2}
which have been developed successfully to formulate quantum
general relativity\footnote{Earlier papers in this direction
include \cite{jac1, super, GSU1, GSU2}.}.

The particular interest in this paper is the form of the boundary
theory, which turns out to be built from the states of a
supersymmetric Chern-Simons theory based on the superalgebra
$Osp(1|2) \oplus Osp(1|2)$. We believe that, when extended to the
${\cal N} =8$ case, these results will be significant for the
understanding of the AdS/CFT conjecture in $3+1$ dimensions. The
extension of our results to the $N=2$ case may also make possible
the detailed comparisons of the string theory and loop quantum
gravity description of boundaries and horizons\cite{linking}.
These questions will be the subject of further papers in this
series.

Along the way, the key idea that we exploit is that supergravity,
as well as general relativity \cite{linking}, can be understood as
a constrained topological quantum field theory. This emerged
already in several papers \cite{GSU1,GSU2}.  This fact is
responsible for the characteristic form of the boundary theory, as
made up of conformal blocks coming from representations of
topological field theory.

For the case in which spacetime has a time like boundary with
finite area, the holographic principle can be stated as follows
\cite{linking,hologr,pluralistic,holo1,holo2,bek}.

1) The hilbert space ${\cal H}_{\cal B}$ for all states of a
quantum gravity theory on the interior of a spacial manifold $\cal
M$ must be decomposable in terms of eigenspaces of an observable
$\hat{A}$ that measures the area of the boundary ${\cal
B}=\partial {\cal M}$. Thus, \f {\cal H}_{\cal B}= \sum_a {\cal
H}_a, \ff where $a$ are the eigenvalues of the area operator
$\hat{A}$. We write the decomposition as a sum because $\hat{A}$
is has been shown to have a discrete spectra, assuming only that
it may be regulated in a way that results in an operator that is
finite and diffeomorphism invariant \cite{sn1}.

2)  The physical state spaces ${\cal H}_{a}$ must have finite
dimension bounded by \f dim({\cal H}_{A}) \leq e^{a \over 4
G\hbar} \label{bek}. \ff

3)  A complete measurement of a state in ${\cal H}_{\cal B}$ must
be possible using only elements of an observable algebra ${\cal
A}_{\cal B}$ associated with measurements that can be made on the
boundary $\cal B$.  The area $\hat{A}$ must be one of these
observables.

This is a powerful clue, because quantum field theories with
finite dimensional Hilbert spaces are not common (even the
harmonic oscillator has an infinite dimensional Hilbert space.)
The only large class of such theories are topological quantum
field theories (TQFT).  A natural strategy for formulating
holographic quantum theories of gravity is then to look for
quantum field theories in $d+1$ dimensions that induce $TQFT$'s on
their $d$ dimensional boundaries.   One advantage of such a
formulation is that it is already in a language which is
background independent and non-perturbative.

As shown in  \cite{linking,hologr}, quantum general relativity is
exactly this kind of theory, as it can be understood to arise from
a topological field theory by the imposition of certain
constraints. We have also found this to be true of $11$
dimensional supergravity, at least at the classical
level\cite{D=11}. The main goal of this paper is to extend that
analysis to quantum supergravity.

In all these cases there is a set of exact physical quantum states
in the bulk, associated with the boundary states
\cite{kodama,superkodama}. Besides being exact descriptions of the
Planck scale structure these states also have semiclassical
interpretations in terms of fluctuations around AntideSitter or
deSitter spacetimes \cite{kodama,chopinlee}.

That such results are possible at all may seem very mysterious,
given that these theories are perturbatively non-renormalizable.
They are possible because these are not just any
non-renormalizable theories; they have special structures, which
are in fact closely related to topological field theories.  In
fact, general relativity and supergravity ( at least for $N=1,2$)
can be formulated as {\it constrained topological field theories.}
This means that the derivative and boundary terms in the action
are the same as in a related topological quantum field theory. The
local degrees of freedom are introduced by a set of local
constraints that do not involve the derivatives of the
gravitational fields, and so do not alter the commutation
relations of the quantum theory.   It is this special structure
that makes possible the holographic formulation of these theories,
as well as a long list of other non-perturbative results, in both
the canonical \cite{reviews} and path-integral \cite{louis,baez}
frameworks.

A holographic formulation has been constructed for the quantum
general relativity.In fact, it has been known for some time that
general relativity in $3+1$ dimensions with a non-vanishing
cosmological constant can be formulated in such a way that it
induces a $2+1$ dimensional $TQFT$ on its finite boundary.  This
theory has exactly the form just described, where the finite
dimensional Hilbert spaces on the finite boundaries of a given
area are built from the conformal blocks of an $SU(2)_q$, $WZW$
conformal field theories on punctured spheres \cite{linking}. The
area of the boundary is given by the sum of the total spins of the
punctures, and the level $k$ is related to the cosmological
constant by $k=6\pi /G^2 \Lambda$ \cite{linking,qdeform}. This new
formulation is treated in \cite{linking}. The basic framework
developed there is based on a representation of general relativity
as a constrained topological field theory ($TFT$). The starting
point is a $TFT$ for an algebra $G$, taken to be $Sp(4)$. This
theory has no local degrees of freedom, but induces degrees of
freedom on finite boundaries which arise from a $2+1$ dimensional
Chern-Simons theory on the boundary.  The local degrees of freedom
are introduced by imposing constraints, which break the gauge
symmetry to a subgroup $H=SU(2)_L \oplus SU(2)_R$. The result is
that the physical degrees of freedom live in the coset $G/H$.  The
degrees of freedom on the boundary are also restricted to those of
a Chern-Simons theory for $H$, but with curvatures constrained by
the degrees of freedom in the coset, which turn out to
parameterize the induced metric in the boundary. Further, the
generators of the gauge transformations for the coset $G/H$ do not
disappear, they instead emerge as the generators of four
dimensional diffeomorphisms.  Extended to the boundary theory,
they define the Hamiltonian of the theory.

In this paper we extend this construction to the N=1 chiral
supergravity. We organize the paper as follows. In section two we
rewrite the N=1 chiral supergravity \cite{jac1} in the formulation
of $Osp(1|4)$ constrained super $BF$ theory. Then in the following
section we give the canonical formalism of the theory. The
boundary formulation of the model which  has finite boundary is
described in section four. In section five we give the quantum
mechanical description of the theory, and show that the space of
boundary states can be constructed from the state space of an
$Osp(1|2)\textstyle\bigoplus Osp(1|2)$ Chern-Simons theory. We
show that as in quantum general relativity,  the Bekenstein bound
is naturally satisfied\cite{linking}. We conclude the paper with a
discussion of future directions.

\section{Supergravity as a constrained topological
field theory}

${\cal N}=1$ supergravity can be written in a chiral formulation
in terms of the pure spin connection \cite{jac1}. Here we'll show
it can be written down directly as a constrained $B\wedge F$
theory by extending the algebra of the connection from $Sp(4)$ to
$Osp(1|4,R)$. This is similar to the way that supergravity was
established as a constrained $BF$ theory by Sano \cite{GSU1} and
Ezawa \cite{GSU2}.

The $Osp(1|4)$ superalgebra is a graded extension of the Sp(4) Lie
algebra. It is generated by bosonic generators $J_{AB},J_{A'B'},$
and $J_{AA'}$ which span the $sp(4)$ algebra and fermionic
generators $Q_{A}$ and $Q_{A'}$, where $(A,A')$ is a pair of
$SU(2)$ indices corresponding to $SU(2)\oplus SU(2)\subset Sp(4)$.
The superalgebra is

\fa
&&[J_{AB},J^{CD}]=\delta_{(A}^{(C}J_{B)}^{D)},\;\;\;\;\;\;\;\;\;\;
[J_{A'B'},J^{C'D'}]=\delta_{(A'}^{(C'}J_{B')}^{D')},\nonumber\\
&&[J_{AA'},J_{BB'}]=\epsilon_{AB}J_{A'B'}+\epsilon_{A'B'}J_{AB},\;\;\;\;
\{Q_{A},Q_{B}\}=\textstyle{\frac{G\sqrt{\Lambda}}{2}}J_{AB},\nonumber\\
&&\{Q_{A'},Q_{B'}\}=\textstyle{\frac{G\sqrt{\Lambda}}{2}}J_{A'B'},\;\;\;\;\;\;\;\;\;\;
\{Q_{A},Q_{A'}\}=\textstyle{\frac{G\sqrt{\Lambda}}{2}}J_{AA'},\nonumber\\
&&[J_{AB},Q_{C}]=\epsilon_{C(A}Q_{B)},\;\;\;\;\;\;\;\;\;\;
 [J_{A'B'},Q_{C'}]=\epsilon_{C'(A'}Q_{B')},\nonumber\\
&&[J_{AA'},Q_{B}]=\epsilon_{AB}Q_{A'},\;\;\;\;\;\;\;\;\;\;
 [J_{AA'},Q_{B'}]=\epsilon_{A'B'}Q_{A},\nonumber\\
&&[J_{AB},Q_{C'}]=0, \;\;\;\;\;\;\;\;\;\;\; [J_{A'B'},Q_{C}]=0,
\ffa where $G$ is the bare gravitational constant and the
$\Lambda$ is cosmology constant.

The supertrace, $Str$, may be defined acting on the generators.
The non-vanishing terms are \fa &&
STr\left\{J_{AB}J^{CD}\right\}:=\delta_{(A}^{(C}\delta_{B)}^{D)}\;\;\;\;\;\;\;\;
STr\left\{J_{A'B'}J^{C'D'}\right\}:=\delta_{(A'}^{(C'}\delta_{B')}^{D')}\;\;\;\;\;\;\;
STr\left\{J_{AA'}J^{BB'}\right\}:=\delta_{A}^{B}\delta_{A'}^{B'}
\nonumber\\ &&
STr\left\{Q_AQ_B\right\}:=\textstyle{\frac{1}{\mu}}\epsilon
_{AB}\;\;\;\;\;\;\;
STr\left\{Q_{A'}Q_{B'}\right\}:=\textstyle{\frac{1}{\mu}}
\epsilon_{A'B'}. \label{str}\ffa The configuration variables of
the theories we will study here are the components of the
connection one form of $Osp(1|4)$. \f \cal A\mit
=A^{AB}J_{AB}+A^{A'B'}J_{A'B'}+\textstyle{\frac{1}{l}}e^{AA'}{J_{AA'}}+
\psi^{A}Q_{A}+\chi^{A'}Q_{A'}, \ff where $l$ is a constant with
the dimension of length.  We see that it includes the vierbein
one-form $e^{AA'}$ and gravitino and anti-gravitino one forms
$\psi^{A}$ and $\chi^{A'}$.

The curvature two form  $\cal F$ is \fa \cal F\mit :&=&d\cal
A\mit+[\cal A\mit, \cal A\mit]\nonumber\\&=&
F_{AB}J^{AB}+F_{A'B'}J^{A'B'}+F_{AA'}J^{AA'}+ F_{A}Q^{A}
+F_{A'}Q^{A'}. \label{F}\ffa The components of the curvature are,
\f F_{AB}=f_{AB}+\textstyle{\frac{1}{l^2}}e_{AA'}\wedge{e_B}^{A'}+
\textstyle{\frac{G\sqrt{\Lambda}}{2l}}\psi_A\wedge\psi_B, \ff \f
F_{A}=d\psi_{A}+A_{A}^{B}\wedge\psi_{B}-\textstyle
{\frac{1}{l}}e_{AA'}\wedge\chi^{A'} =
D\psi_{A}-\textstyle{\frac{1}{l}}e_{AA'}\wedge\chi^{A'} \equiv
f_A-\textstyle{\frac{1}{l}}e_{AA'}\wedge\chi^{A'}, \ff \f
F_{A'B'}=f_{A'B'}+\textstyle{\frac{1}{l^2}}e_{A'A}\wedge
e_{B'}^{A}+
\textstyle{\frac{G\sqrt{\Lambda}}{2l}}\chi_{A'}\wedge\chi_{B'},
\ff \f F_{A'}=d\chi_{A'}+{A_{A'}}^{B'}
\wedge\chi_{B'}-\textstyle{\frac{1}{l}}e_{A'A}\wedge\psi^{A}
=\bar{D}\chi_{A'}-\textstyle{\frac{1}{l}}e_{A'A}\wedge\psi^{A}
\equiv f_{A'}-\textstyle{\frac{1}{l}}e_{A'A}\wedge\psi^{A}, \ff \f
F_{AA'}=de_{AA'}+A_{A}^{B}\wedge e_{BA'}+A_{A'}^{B'}\wedge
e_{AB'}+\textstyle{\frac{G\sqrt{\Lambda}}{2l}}\psi_{A}\wedge\chi_{A'}.
\ff

To construct the action we will need a Lie algebra valued two form
$\cal B$, whose components are labeled. \f \cal B\mit
:=B_{AB}J^{AB}+B_{A'B'}J^{A'B'}+B_{AA'}J^{AA'}+\mu B_{A}Q^{A}+\mu
B_{A'}Q^{A'}. \label{B} \ff

We can now write the action for the $Osp(1|4)$ $BF$ theory.

\f -i\cal I\mit =-i\int_{M}STr\{\textstyle{\frac{1}{g^2}}\cal
B\wedge F\mit-\textstyle\frac{e^2}{2}\cal B\wedge B\}. \ff Here
$g$,$e$ are dimensionless constants. Note that $\mu$ appearing in
($\ref{str}$) and ($\ref{B}$) is another dimensionless constant.

The super-$BF$ theory is a topological quantum field theory in
that it has no local degrees of freedom We now follow the method
introduced in \cite{hologr} and construct the action for
supergravity by constraining the fields of the topological
super-$BF$ theory.  To do this we break some of the gauge
invariance by imposing local, non-derivative constraints.  The
supergravity action is \f -i\cal I\mit^{initial}=-i\cal
I\mit^{SBF}-i\cal I\mit^{Constraint}, \label{act} \ff where \f
I\mit^{Constraint}= \int_{\cal{M}}
\lambda^{AB}(\textstyle{\frac{1}{l^2}}e_{AA'}\wedge
{e_B}^{A'}-B_{AB})+\lambda^{A}(\textstyle{\frac{1}{l}}e_{AA'}\wedge
{\chi}^{A'}-B_{A}). \ff In components this action is \fa
-i\cal{I}\mit^{initial} &
=&\int_{\cal{M}}\textstyle{\frac{1}{g^2}}(B_{AB}\wedge
F^{AB}+\frac{1}{l} B_{A}\wedge
F^{A})-\textstyle{\frac{e^2}{2}}(B_{AB}\wedge
B^{AB}+\frac{\mu}{l}B_{A}\wedge B^{A})\nonumber\\ && +
\textstyle{\frac{1}{g^2}}(B_{A'B'}\wedge F^{A'B'}+\frac{1}{l}
B_{A'}\wedge F^{A'})-\textstyle{\frac{e^2}{2}}(B_{A'B'}\wedge
B^{A'B'}+\frac{\mu}{l}B_{A'}\wedge B^{A'})\nonumber\\ &&
+\textstyle{1 \over g^{2}}(B_{AA'}\wedge F^{AA'} -\textstyle
{e^{2}\over 2} B^{AA'}\wedge B_{AA'})\nonumber\\ &&
+\lambda^{AB}(\textstyle{\frac{1}{l^2}}e_{AA'}\wedge
{e_B}^{A'}-B_{AB})+\lambda^{A}(\textstyle{\frac{1}{l}}e_{AA'}\wedge
{\chi}^{A'}-B_{A})\nonumber\\ &&
+\lambda^{A'B'}(\textstyle{\frac{1}{l^2}}e_{AA'}\wedge
{e_B'}^{A}-B_{A'B'})+\lambda^{A'}(\textstyle{\frac{1}{l}}e_{AA'}\wedge
{\psi}^{A}-B_{A'}). \label{actiontotal} \ffa Note that the field
equation for $B^{AA'}$ yields, \f F_{AA'}=B_{AA'}. \ff Thus
$B_{AA'}$ carries the information as to the torsion, so that \f
B_{AA'} = \bigtriangledown
e_{AA'}+\textstyle{\frac{G\sqrt{\Lambda}}{2l}}
\psi_{A}\wedge\chi_{A'}. \label{torsion} \ff The other field
equations will eventually set $B^{AA'}$ to zero, hence this part
of the action is redundant and can be dropped. This simplifies the
resulting canonical analysis, but does not affect the field
equations.  Once this is done the action splits into left and
right handed pieces \f -i\cal{I}\mit^{SG}=
-i\cal{I}\mit^{SG}_{L}+-i\cal{I}\mit^{SG}_{R}, \ff where \fa
-i\cal{I}\mit^{SG}_{L} &
=&\int_{\cal{M}}\textstyle{\frac{1}{g^2}}(B_{AB}\wedge
F^{AB}+\frac{1}{l} B_{A}\wedge
F^{A})-\textstyle{\frac{e^2}{2}}(B_{AB}\wedge
B^{AB}+\frac{\mu^2}{l}B_{A}\wedge B^{A})\nonumber\\ &
&+\lambda^{AB}(\textstyle{\frac{1}{l^2}}e_{AA {}^\prime}\wedge
  {e_B}^{A
{}^\prime}-B_{AB})+\lambda^{A}(\textstyle{\frac{1}{l}}
  e_{AA {}^\prime}\wedge
{\chi}^{A {}^\prime}-B_{A}). \label{action0} \ffa The right handed
piece is formally the conjugate of the left handed piece that
results from switching primed and unprimed indices. We will see
shortly how the reality conditions arise, whose effect will be to
require that they are complex conjugates of each other.

To see that (\ref{action0}) is an action for $N=1$ supergravity we
proceed to solve the constraint for $B_{AB}$ and $B_A$

\f B_{AB}=\textstyle{\frac{1}{l^2}}e_{AA'}\wedge {e_B}^{A'},\ff \f
B_{A}=\textstyle{\frac{1}{l}}e_{AA'}\wedge {\chi}^{A'}.\ff
Substituting the solutions to $(\ref{action0})$, we find the
action taking the form \fa
-i\cal{I}\mit_{L}&=&\int_{\cal{M}}\textstyle{\frac{1}{g^2l^2}}(e_{AA'}\wedge
   {e_B}^{A'} \wedge f^{AB}+e_{AA'}\wedge
{\chi}^{A'}\wedge D\psi^{A})\nonumber\\
&&-\textstyle{\frac{1}{l^4}(\frac{1}{g^2}-\frac{e^2}{2})}(e_{AA'}\wedge
{e_B}^{A'}
   \wedge
e^{AB'}\wedge{e^B}_{B'})+\textstyle{\frac{\sqrt\Lambda}{2l}}(e_{AA'}\wedge
  {e_B}^{A'}\wedge\psi^A\wedge\psi^B)\nonumber\\
&&-\frac{1}{l}(\textstyle{\frac{e^2\mu}{2l^2}}+\textstyle{\frac{1}{g^2l^2}})(e_{AA'}
  \wedge\chi^{A'}\wedge {e^A}_{B'}\wedge\chi^{B'}).
\ffa If we define the dimensionless constants as follows \f
G:=g^2l^2; \ \ \
\Lambda:=\textstyle{\frac{6}{l^4}(\frac{1}{g^2}-\frac{e^2}{2})}; \
\ \ \mu:=\textstyle{\frac{G\sqrt{\Lambda}-6}{3g^2e^2}}\ff Then:
\fa -i\cal{I}\mit_{L}
&=&\int_{\cal{M}}\textstyle{\frac{1}{G}}(e_{AA'}\wedge
{e_B}^{A'}\wedge f^{AB} +e_{AA'}\wedge\chi^{A'}\wedge
D\psi^{A})\nonumber\\ &&
-\textstyle{\frac{\Lambda}{6}}(e_{AA'}\wedge {e_B}^{A'}\wedge
e^{AB'}\wedge e_{B'}^B)\nonumber\\
&&+\textstyle{\frac{\sqrt\Lambda}{2l}}(e_{AA'}\wedge
{e_B}^{A'}\wedge\psi^A\wedge\psi^B)-
\textstyle{\frac{\sqrt{\Lambda}}{6l}}(e_{AA'}\wedge
\chi^{A'}\wedge {e^A}_{B'}\wedge\chi^{B'}). \ffa This is the same
action as the CDJ formalism \cite{CDJ} after we solve the
constraint equations associated to the Lagrangian multipliers
$\phi_{ABCD}$ and $\kappa_{ABC}$ \cite{jac1}. We may note that the
cosmological constant is zero if $e^2g^2=2$, at which point
$\mu=-1$.

\section{The Canonical Formalism of N=1 SCBF theory}

We now study the canonical formalism for $N=1$ supergravity based
on the fields just introduced. Our main goal is to understand
those issues which arise in the supergravity case. These mainly
have to do with how the anticommutation relations between the left
and right handed supersymmetry constraints arise and how they come
to close on the spatial diffeomorphism and hamiltonian
constraints.

In this section we ignore the possibility of boundary terms. Many
of the expressions will later be modified by the presence of
boundary terms.

\subsection{The $3+1$ decomposition of the action.}

In the last section we see that $N=1$ supergravity can be written
as a constrained topological field theory based on the supergroup
$Osp(1|4)$. The total action can be expressed in terms of the
component fields

\fa -i\cal{I}\mit
&=&-i\int_{M}dx^4\{\textstyle{\frac{1}{g^2}}(B_{AB}\wedge
F^{AB}+\frac{1}{l} B_{A}\wedge
F^{A})-\textstyle{\frac{e^2}{2}}(B_{AB}\wedge
B^{AB}+\frac{\mu^2}{l}B_{A}\wedge B^{A})\nonumber\\
&&+\lambda^{AB}(\textstyle{\frac{1}{l^2}}e_{AA'}\wedge{e_B}^{A'}-B_{AB})
+\lambda^{A}(\textstyle{\frac{1}{l}}e_{AA'}\wedge
{\chi}^{A'}-B_{A})\nonumber\\
&&-\textstyle{\frac{1}{g^2}}(B_{A'B'}\wedge F^{A'B'}+
\frac{1}{l}B_{A'}\wedge
F^{A'})+\textstyle{\frac{e^2}{2}}(B_{A'B'}\wedge
B^{A'B'}+\frac{\mu^2}{l}B_{A'}\wedge B^{A'})\nonumber\\
&&-\lambda^{A'B'}(\textstyle{\frac{1}{l^2}}e_{A'A}\wedge{e_B'}^{A}-B_{A'B'})
-\lambda^{A'}(\textstyle{\frac{1}{l}}e_{A'A}\wedge{\psi}^{A}-B_{A'})\}.
\ffa We proceed with the 3+1 decomposition of the action. We
assume the spacetime has hyperbolic structure and has the topology
of $R\times \Sigma$. We then express the action in terms of space
and time independently \fa -i\cal{I}\mit &=&-i\int
dt\int_{\Sigma}dx^3\epsilon^{abc}\{
\frac{1}{g^2}({B_{ab}}^{AB}\dot{A}_{cAB}+\frac{1}{l}{B_{ab}}^{A}\dot{\psi}_{cA}-
{B_{ab}}^{A'B'}\dot{A}_{cA'B'}-\frac{1}{l}{B_{ab}}^{A'}\dot{\chi}_{cA'})\nonumber\\
&&+\frac{1}{g^2}A_0^{AB}(D_{a}B_{bcAB}+\frac{1}{l}B_{bcA}\psi_{aB})
-\frac{1}{g^2}A_0^{A'B'}
(\bar{D}_{a}B_{bcA'B'}+\frac{1}{l}B_{bcA'}\chi_{aB'})\nonumber\\
&&+\frac{1}{l}{\psi_0}^{A}\left[\textstyle{\frac{1}{g^2}}(D_{a}B_{bcA}+\textstyle{G\sqrt{\Lambda}}B_{bcAB}\psi_{a}^{B})+
{\lambda^{A'}}_{ab}e_{cA'A}-\textstyle{\frac{1}{g^2l}}B_{ab}^{A'}e_{cA'A}\right]\nonumber\\
&&-\frac{1}{l}{\chi_0}^{A'}\left[\textstyle{\frac{1}{g^2}}(\bar{D}_{a}B_{bcA'}+\textstyle{G\sqrt{\Lambda}}
B_{bcA'B'}\chi_{a}^{B'})+
{\lambda^{A}}_{ab}e_{cAA'}-\textstyle{\frac{1}{g^2l}}B_{ab}^{A}e_{cAA'}\right]\nonumber\\
&&+e_{AA'0}[\textstyle{\frac{2}{g^2l^2}}B_{ab}^{AB}{e_{cB}}^{A'}+\textstyle{\frac{2}{l^2}}
\lambda_{ab}^{AB}{e_{cB}}^{A'}+\textstyle{\frac{1}{l}}
\lambda_{ab}^{A}\chi_{c}^{A'}-\textstyle{\frac{1}{g^2l^2}}B_{ab}^{A}\chi_{c}^{A'}\nonumber\\
&&-\textstyle{\frac{2}{g^2l^2}}B_{ab}^{A'B'}e_{cB'}^A
-\textstyle{\frac{2}{l^2}}\lambda_{ab}^{A'B'}{e_{cB'}}^{A}-\textstyle{\frac{1}{l}}
\lambda_{ab}^{A'}\psi_{c}^{A}+\textstyle{\frac{1}{g^2l^2}}B_{ab}^{A'}\psi_{c}^{A}]\nonumber\\
&&+B_{0a}^{AB}\left[\textstyle{\frac{1}{g^2}}f_{bcAB}+\textstyle{\frac{1}{g^{2}l^{2}}}e_{bA}^{A'}e_{cBA'}+
\textstyle{\frac{1}{g^2}\frac{G\sqrt{\Lambda}}{2l}}\psi_{bA}\psi_{cB}-e^{2}B_{bcAB}-
\lambda_{bcAB}\right]\nonumber\\
&&-B_{0a}^{A'B'}\left[\textstyle{\frac{1}{g^2}}f_{bcA'B'}+\textstyle{\frac{1}{g^2l^2}}{e_{bA'}}^{A}e_{cB'A}+
\textstyle{\frac{1}{g^2}\frac{G\sqrt{\Lambda}}{2l}}\chi_{bA'}\chi_{cB'}-e^{2}B_{bcA'B'}-
\lambda_{bcA'B'}\right]\nonumber\\
&&+\frac{1}{l}B_{0a}^{A}\left[\textstyle{\frac{1}{g^2}}f_{bcA}-\textstyle{\frac{1}{g^2l}}e_{bAA'}\chi_{c}^{A'}
-e^{2}\mu^{2}B_{bcA}-l\lambda_{bcA}\right]\nonumber\\
&&-\frac{1}{l}B_{0a}^{A'}\left[\textstyle{\frac{1}{g^2}}f_{bcA'}-\textstyle{\frac{1}{g^2l}}B_{bA'A}\psi_{c}^{A}
-e^{2}\mu^{2}B_{bcA'}-l\lambda_{bcA'}\right]\nonumber\\
&&+\lambda_{0a}^{AB}\left[\textstyle{\frac{1}{l^2}}e_{bA}^{A'}e_{cBA'}-B_{bcAB}\right]-
\lambda_{0a}^{A'B'}\left[\textstyle{\frac{1}{l^2}}e_{bA'}^{A}e_{cB'A}-B_{bcA'B'}\right]\nonumber\\
&&+\lambda_{0a}^{A}\left[\textstyle{\frac{1}{l}}e_{bA}^{A'}\chi_{cA'}-B_{bcA}\right]-
\lambda_{0a}^{A'}\left[\textstyle{\frac{1}{l}}e_{bA'}^{A}\psi_{cA}-B_{bcA'}\right]\}.
\label{action} \ffa Here $0$ is the  time-like index and
$a,b,c(=1,2,3)$ are space-like indices. From (\ref{action}), we
define the non-vanishing momenta for the forms
$A_{AB},A_{A'B'},\psi_{A},\chi_{A'}$ as\footnote{These will
receive corrections when we introduce the boundary terms.} \f
\pi_{AB}^{a}:=\textstyle{\frac{-i}{g^2}}\epsilon^{abc}B_{bcAB}\;\;\;\;\;
\pi_{A'B'}^{a}:=\textstyle{\frac{i}{g^2}}\epsilon^{abc}B_{bcA'B'},\ff
\f
\pi_{A}^{a}:=\textstyle{\frac{-i}{g^2}}\epsilon^{abc}B_{bcA}\;\;\;\;\;
\pi_{A'}^{a}:=\textstyle{\frac{i}{g^2}}\epsilon^{abc}B_{bcA'}. \ff

The other momenta for the forms $B,\lambda,$and $e_{0A'B'}$
vanish. We then rewrite the action as \fa -i\cal{I}\mit &=&\int
dt\int_{\Sigma}dx^3\{
(\pi_{AB}^{a}\dot{A}_{a}^{AB}+\frac{1}{l}\pi_{A}^{a}\dot\psi_{a}^{A}+
\pi_{A'B'}^{a}\dot{A}_{a}^{A'B'}+\frac{1}{l}\pi_{A'}^{a}\dot\chi_{a}^{A'})
\nonumber\\
&&+A_0^{AB}(D_{a}\pi_{AB}^{a}+\frac{1}{l}\pi_{A}^{a}\psi_{aB})
+A_0^{A'B'}
(\bar{D}_{a}\pi_{A'B'}^{a}+\frac{1}{l}\pi_{A'}^{a}\chi_{aB'})\nonumber\\
&&+\frac{1}{l}{\psi_0}^{A}\left[(D_{a}\pi_{A}^{a}+\textstyle{G\sqrt{\Lambda}}
\pi_{AB}^{a}\psi_{a}^{B})
-i\epsilon^{abc}{\lambda^{A'}}_{ab}e_{cA'A}+
\textstyle{\frac{1}{l}}\pi^{aA'}e_{aA'A}\right]\nonumber\\
&&-\frac{1}{l}{\chi_0}^{A'}\left[(\bar{D}_{a}\pi_{A'}^{a}+
\textstyle{G\sqrt{\Lambda}}
\pi_{A'B'}^{a}\chi_{a}^{B'})+i\epsilon^{abc}{\lambda^{A}}_{ab}e_{cAA'}
+\textstyle{\frac{1}{l}}\pi^{aA}e_{aAA'}\right]\nonumber\\
&&+e_{AA'0}[\textstyle{\frac{2}{l^2}}\pi^{aAB}e_{aB}^{A'}-\textstyle{\frac{1}{l^2}}\pi^{aA}\chi_{a}^{A'}
-\textstyle{\frac{2i}{l^2}}\epsilon^{abc}\lambda_{ab}^{AB}e_{cB}^{A'}-\textstyle{\frac{i}{l}}\epsilon^{abc}
\lambda_{ab}^{A}\chi_{c}^{A'}\nonumber\\
&&+\textstyle{\frac{2}{l^2}}\pi^{aA'B'}e_{aB'}^A-\textstyle{\frac{1}{l^2}}\pi^{aA'}\psi_a^A
+\textstyle{\frac{2i}{l^2}}\epsilon^{abc}\lambda_{ab}^{A'B'}e_{cB'}^{A}+\textstyle{\frac{i}{l}}\epsilon^{abc}
\lambda_{ab}^{A'}\psi_{c}^{A}]\nonumber\\
&&-iB_{0a}^{AB}\left[\textstyle{\frac{1}{g^2}}\epsilon^{abc}f_{bcAB}+
\textstyle{\frac{1}{g^2l^2}}\epsilon^{abc}e_{bA}^{A'}e_{cBA'}+
\textstyle{\frac{1}{g^2}\frac{G\sqrt{\Lambda}}{2l}}\epsilon^{abc}\psi_{Ab}\psi_{Bc}-ie^{2}g^{2}\pi_{AB}^{a}-
\epsilon^{abc}\lambda_{bcAB}\right]\nonumber\\
&&+iB_{0a}^{A'B'}\left[\textstyle{\frac{1}{g^2}}\epsilon^{abc}f_{bcA'B'}+
\textstyle{\frac{1}{g^2l^2}}\epsilon^{abc}e_{bA'}^{A}e_{cB'A}+
\textstyle{\frac{1}{g^2}\frac{G\sqrt{\Lambda}}{2l}}\epsilon^{abc}\chi_{A'b}\chi_{B'c}+ie^{2}g^{2}\pi_{A'B'}{a}-
\epsilon^{abc}\lambda_{bcA'B'}\right]\nonumber\\
&&-\frac{i}{l}B_{0a}^{A}\left[\textstyle{\frac{1}{g^2}}\epsilon^{abc}f_{bcA}
-\textstyle{\frac{1}{g^2l}}\epsilon^{abc}e_{bAA'}\chi_{c}^{A'}-ie^{2}g^{2}\mu^{2}\pi_{A}^{a}-
l\epsilon^{abc}\lambda_{bcA}\right]\nonumber\\
&&+\frac{i}{l}B_{0a}^{A'}\left[\textstyle{\frac{1}{g^2}}\epsilon^{abc}f_{bcA'}
-\textstyle{\frac{1}{g^2l}}\epsilon^{abc}e_{bA'A}\psi_{c}^{A}+ie^{2}g^{2}\mu^{2}\pi_{A'}^{a}-
l\epsilon^{abc}\lambda_{bcA'}\right]\nonumber\\
&&-i\lambda_{0a}^{AB}\left[\textstyle{\frac{1}{l^2}}\epsilon^{abc}e_{bA}^{A'}e_{cBA'}-ig^2\pi_{AB}^{a}\right]+i
\lambda_{0a}^{A'B'}\left[\textstyle{\frac{1}{l^2}}
\epsilon^{abc}e_{bA'}^{A}e_{cB'A}+ig^2\pi_{A'B'}^{a}\right]\nonumber\\
&&-i\lambda_{0a}^{A}\left[\textstyle{\frac{1}{l}}\epsilon^{abc}e_{bA}^{A'}\chi_{cA'}-ig^2\pi_{A}^{a}\right]+i
\lambda_{0a}^{A'}\left[\textstyle{\frac{1}{l}}\epsilon^{abc}e_{bA'}^{A}\psi_{cA}+ig^2\pi_{A'}^{a}\right]\}.
\label{action2} \ffa

\subsection{The primary constraints}

We first consider those  constraints associated with Lagrange
multipliers $\lambda_{0}$.  The solutions to these equations will
help to simplify the other constraints. To break the topological
gauge symmetry of the $BF$ theory, and in the process introduce
the local degrees of freedom, additional constraints have been
introduced in (\ref{action}). Their canonical form can be read off
of (\ref{action2}), yielding \f
J^{a}_{AB}:=\textstyle{\frac{1}{l^2}}
\epsilon^{abc}e_{bA}^{A'}e_{cBA'}-ig^2\pi_{AB}^{a}=0, \label{jAB}
\ff \f J^{a}_{A'B'}:=\textstyle{\frac{1}{l^2}}
\epsilon^{abc}e_{bA'}^{A}e_{cB'A}+ig^2\pi_{A'B'}^{a}=0,
\label{jA'B'} \ff \f J^{a}_{A}:=\textstyle{\frac{1}{l}}
\epsilon^{abc}e_{bA}^{A'}\chi_{cA'}-ig^2\pi^{a}_{A}=0, \label{JA}
\ff \f J^{a}_{A'}:=\textstyle{\frac{1}{l}}
\epsilon^{abc}e_{bA'}^{A}\psi_{cA}+ig^2\pi^{a}_{A'}=0. \label{JA'}
\ff These equations set the $\pi^{aA'B'},\pi^{aAB},\pi^{aA}$ and
$\pi^{aA'}$ to be the duals of the two forms constructed by the
frame fields and the spinor fields.

We first discuss the solution to the bosonic constraints
(\ref{jAB}) and (\ref{jA'B'}). These are the same as in general
relativity and we refer the reader to \cite{hologr} for more
discussion of the following points.
 We introduce the quantities $N_{AA'}$ which is
defined as \f N_{AA'}=t^{\mu}e_{\mu AA'},\ff where $t^{\mu}$ is
the timelike unit normal satisfying
 $t^{\mu}t_{\mu}=-1$. Then we can express
$e_{a}^{AA'}$,
 in terms of $N_B^{A'}$ and either  $\pi^{aAB}$ or
$\pi^{aA'B'}$ \fa
e_a^{AA'}&=&\textstyle{\frac{1}{\sqrt{h}}}\epsilon_{abc}
\pi^{bBC}\pi^{cA}_{C}N_B^{A'},\label{d1}\\
e_a^{A'A}&=&\textstyle{\frac{1}{\sqrt{h}}}\epsilon_{abc}
\pi^{bB'C'}\pi^{cA'}_{C'}N_{B'}^A\label{d2},\ffa where $h$ is the
determinant of the spatial metric $h_{ab}$.

Notice that there is a secondary constraint, which is

\f R_{a}^{AA'}=\textstyle{\frac{1}{\sqrt{h}}}\epsilon_{abc}
\pi^{bBC}\pi^{cA}_{C}N_B^{A'}-\textstyle{\frac{1}{\sqrt{h}}}\epsilon_{abc}
\pi^{bB'C'}\pi^{cA'}_{C'}N_{B'}^A.\label{RaAA'} \ff This expresses
the reality conditions. We will later use a consequence of this,
which expresses the idea that the areas of surfaces defined from
the left handed fields are equal to the areas defined from the
right handed fields\cite{hologr}.

We now come to the treatment of the fermion variables.
 There is a
difficulty which arises from  equations (\ref{JA}) and
(\ref{JA'}). These tell us that the Poisson antibracket $\{\psi_A,
\chi_{A'}\}$ is non-zero.  Hence  the configuration space can't
contain both fermionic fields $\psi_A$ and $\chi_{A'}$ and we
cannot construct a quantum theory in terms of simultaneous
eigenstates of $\psi_A$ and $\chi_{A'}$. This is a well known
problem in fermionic theories, there are two ways of handling it.

\subsection{Treatment of the fermionic constraints:
method 1:}

We break the left-right symmetry immediately and choose,
arbitrarily to diagonalize one of the fermion fields, say
$\psi_A$, while treating the other, $\chi_{A'}$, as a momentum
field. This means that we will choose a representation such that
\f \hat{\psi}^A|\Gamma\rangle=\psi^A|\Gamma\rangle, \ff in which
the action of $\chi_{A'}$ will be  \f
\hat{\chi}^{A'}|\Gamma\rangle=e^{AA'}
\textstyle{\frac{\delta}{\delta\psi^A}}|\Gamma\rangle.\ff Here for
convenience we'd like to choose the conjugate pairs
$(A_a^{AB},\pi^a_{AB}), (A_a^{A'B'}, \pi^a_{A'B'})$, and
$(\psi_a^{A}, \pi^a_A)$ for the phase space, and the Poisson
brackets are \fa \{\pi^{aAB}(x), A_{bCD}(y)\}&=&
\delta^a_b\delta^A_{(C}\delta^{B}_{D)}\delta^3(x,y),\nonumber\\
\{\pi^{aA'B'}(x), A_{bC'D'}(y)\}&=&\delta^a_b\delta^{A'}_{(C'}
\delta^{B'}_{D')}\delta^3(x,y),\nonumber\\ \{\pi^{aA}(x),
\psi_{bB}(y)\}&=&-\delta^a_b\delta^A_B\delta^3(x,y).\ffa with the
rest being zero. We need to solve for the other variables
$e_{a}^{AA'},\pi^{aA'},\chi_a^{A'}$ in terms of the canonical
momenta from equations $(\ref{d1})$-$(\ref{d4})$. To do this we
solve the remaining primary constraints (\ref{JA}) and (\ref{JA'})
to find, \fa
\chi_c^{A'}&=&\textstyle{\frac{1}{\sqrt{h}}\frac{1}{l}}
\epsilon_{abc}\pi^{aAB}\pi^{b}_{A}N_{B}^{A'}, \label{d3}\\
\pi^{a}_{A'}&=&\textstyle{\frac{i}{g^2l}}
\epsilon^{abc}e_{bA'}^{A}\psi_{cA}. \label{d4} \ffa

Proceeding from here we quickly reach the form of canonical
supergravity discussed already by Jacobson in \cite{jac1}.

\subsection{Treatment of the fermionic constraints:
method 2:}

The second method is to keep left-right symmetry at the cost of
keeping in the theory both sets of fermionic variables, along with
the constraints that imply their redundancy.  This will be
convenient for the study of the boundary theory as well as the
quantization.

To do this we find the secondary constraints which are analogous
to (\ref{RaAA'}) which impose the relations between the left and
right fermionic variables.  Taking linear combinations of
(\ref{JA}) and (\ref{JA'}) we find that \f \psi_{a}^{A}
J^{a}_{A}:=\textstyle{\frac{1}{l}} \epsilon^{abc}\psi_{a}^{A}
e_{bA}^{A'}\chi_{cA'}-ig^2\psi_{a}^{A}\pi^{a}_{A}=0, \label{psiJ}
\ff \f \chi_{a}^{A'}J^{a}_{A'}:=\textstyle{\frac{1}{l}}
\epsilon^{abc}\chi_{a}^{A'} e_{bA'}^{A}\psi_{cA}+ig^2\chi_{a}^{A'}
\pi^{a}_{A'}=0. \label{chiJ} \ff By adding and subtracting we get
an equivalent set of constraints \f R=
\psi_{a}^{A}\pi^{a}_{A}-\chi_{a}^{A'} \pi^{a}_{A'}=0, \label{R}
\ff \f S^{0}= \epsilon^{abc}\chi_{a}^{A'} e_{bA'}^{A}\psi_{cA}=0.
\ff eliminating $e_{bA'}^{A}$ we have \f S= \chi_{a}^{A'}
N_{A'}^{B }\pi^{[a}_{BC}\pi^{b]CA}\psi_{bA}=0. \label{S} \ff

\subsection{Elimination of the lagrange multipliers
$\lambda$.}

We will want to eliminate the $\lambda_{AB}$ and $\lambda_{A'B'}$
from the canonical theory.  This can be done by solving the
constraints that follow from the vanishing of the canonical
momenta for $B_{AB},B_{A'B'},B_A,B_{A'}$.  These are \f
I^{a}_{AB}:=\textstyle{\frac{1}{g^2}}\epsilon^{abc}f_{bcAB}+
\textstyle{\frac{1}{g^2l^2}}\epsilon^{abc}e_{bA}^{A'}e_{cBA'}+
\textstyle{\frac{1}{g^2}\frac{G\sqrt{\Lambda}}{2l}}
\epsilon^{abc}\psi_{Ab}\psi_{Bc}-ie^{2}g^{2}\pi_{AB}^{a}-
\epsilon^{abc}\lambda_{bcAB}=0, \ff \f
I^{a}_{A'B'}:=\textstyle{\frac{1}{g^2}}\epsilon^{abc}f_{bcA'B'}+
\textstyle{\frac{1}{g^2l^2}}\epsilon^{abc}e_{bA'}^{A}e_{cB'A}+
\textstyle{\frac{1}{g^2}\frac{G\sqrt{\Lambda}}{2l}}
\epsilon^{abc}\chi_{A'b}\chi_{B'c}+ie^{2}g^{2}\pi_{A'B'}^{a}-
\epsilon^{abc}\lambda_{bcA'B'}=0, \ff \f
I^{a}_{A}:=\textstyle{\frac{1}{g^2}}\epsilon^{abc}f_{bcA}
-\textstyle{\frac{1}{g^2l}}\epsilon^{abc}e_{bAA'}\chi_{c}^{A'}-ie^{2}g^{2}\mu^{2}\pi_{A}^{a}-
l\epsilon^{abc}\lambda_{bcA}=0,\ff \f
I^{a}_{A'}:=\textstyle{\frac{1}{g^2}}\epsilon^{abc}f_{bcA'}
-\textstyle{\frac{1}{g^2l}}\epsilon^{abc}e_{bA'A}\psi_{c}^{A}+ie^{2}g^{2}\mu^{2}\pi_{A'}^{a}-
l\epsilon^{abc}\lambda_{bcA'}=0. \ff These are second class
constraints and can be solved to eliminate $\lambda_{bcAB}$, $
\lambda_{bcA'B'}$, $\lambda_{bcA},$ and $\lambda_{bcA'}$
 in terms of the other variables:
\f
\epsilon^{abc}\lambda_{bcAB}=\textstyle{\frac{1}{g^2}}\epsilon^{abc}f_{bcAB}+
i\pi^a_{AB}+ \textstyle{\frac{1}{g^2}\frac{G\sqrt{\Lambda}}{2l}}
\epsilon^{abc}\psi_{bA}\psi_{cB}-ie^{2}g^{2}\pi_{AB}^{a},\label{l1}\ff
\f
\epsilon^{abc}\lambda_{bcA'B'}=\textstyle{\frac{1}{g^2}}\epsilon^{abc}f_{bcA'B'}-
i\pi^a_{A'B'}+ \textstyle{\frac{1}{g^2}\frac{G\sqrt{\Lambda}}{2l}}
\epsilon^{abc}\chi_{A'b}\chi_{B'c}+ie^{2}g^{2}\pi_{A'B'}^{a},\label{l2}\ff
\f
l\epsilon^{abc}\lambda_{bcA}=\textstyle{\frac{1}{g^2}}\epsilon^{abc}f_{bcA}-i\pi_{A}^{a}
-ie^{2}g^{2}\mu^{2}\pi_{A}^{a},\label{l3}\ff \f
l\epsilon^{abc}\lambda_{bcA'}=\textstyle{\frac{1}{g^2}}\epsilon^{abc}f_{bcA'}+i\pi_{A'}^{a}
+ie^{2}g^{2}\mu^{2}\pi_{A'}^{a}.\label{l4}\ff

\subsection{The Gauss and supersymmetry constraints}

Now we come to the constraints that impose the
$Osp(1|2)_{L}\otimes Osp(1|2)_{R}$ local gauge symmetry.  These
are \fa G_{AB}& :=&
D_{a}\pi_{AB}^{a}+\frac{1}{l}\pi_{A}^{a}\psi_{aB}=0,
\\
G^L_{A}&:=&
D_{a}\pi_{A}^{a}+\textstyle{G\sqrt{\Lambda}}\pi_{AB}^{a}\psi_{a}^{B}
+\textstyle{\frac{1}{l}}\pi^{aA'}e_{aA'A}-i\epsilon^{abc}
{\lambda^{A'}}_{ab}e_{cA'A}\nonumber\\
&=&D_{a}\pi_{A}^{a}+\textstyle{G\sqrt{\Lambda}}
\pi_{AB}^{a}\psi_{a}^{B}+(\textstyle{\frac{-i}{g^2l}}
\epsilon^{abc}f_{bc}^{A'}+\textstyle{\frac{G\sqrt{\Lambda}}{3l}}\pi^{aA'})e_{aA'A}
=0, \label{super}\\
G_{A'B'}&:=&\bar{D}_{a}\pi_{A'B'}^{a}+\frac{1}{l}\pi_{A'}^{a}\chi_{aB'}=0,
\\
G^R_{A'}&:=&\bar{D}_{a}\pi_{A'}^{a}+\textstyle{G\sqrt{\Lambda}}
\pi_{A'B'}^{a}\chi_{a}^{B'}
+\textstyle{\frac{1}{l}}\pi^{aA}e_{aAA'}+i\epsilon^{abc}
{\lambda^{A}}_{ab}e_{cAA'}\nonumber\\&=&\bar{D}_{a}\pi_{A'}^{a}+\textstyle{G\sqrt{\Lambda}}
\pi_{A'B'}^{a}\chi_{a}^{B'} +(\textstyle{\frac{i}{g^2l}}
\epsilon^{abc}f_{bc}^{A}+\textstyle{\frac{G\sqrt{\Lambda}}{3l}}\pi^{aA})e_{aAA'}=0.\label{super2}
\ffa where $(\ref{l3})$ $(\ref{l4})$ are used to cancel the
multipliers $\lambda^A_{ab}$ and $\lambda^{A'}_{ab}$ out of the
equations $(\ref{super})$ and $(\ref{super2})$. Here $G_{AB}$ and
$G_{A'B'}$ are nothing but the ordinary $SU(2)_{L}\oplus SU(2)_{R}
$ Gauss law constraints, and $G^L_{A},G^R_{A'}$ are the
left-handed and right-handed supersymmetry constraints
respectively, which are  different from those of $\it chiral$
supergravity due to the appearance of $e_{AA'}$, which mix the
left-handed supersymmetry constraints $(L_A)$ and right-handed
ones $(R_{A'})$ in chiral supergravities together in the form: \f
G^L_A=L_A+\bar{R}_{A}\;\;\;\;\;\;\;\;
G^R_{A'}=\bar{L}_{A'}+R_{A'}, \ff where \fa
L_A:&=&D_{a}\pi_{A}^{a}-\textstyle{G\sqrt{\Lambda}}
\pi_{AB}^{a}\psi_{a}^{B},\\
R_{A'}:&=&(\textstyle{\frac{-i}{g^2l}}\epsilon^{abc}f_{bc}^{A}
+\textstyle{\frac{G\sqrt{\Lambda}}{3l}}\pi^{aA})e_{aAA'},
 \ffa
and $R_{A'}$ seems not to be conjugate of $L_A$\footnote{This is
the characteristic of chiral supergravity, see \cite{jac1}.}.
However, using $(\ref{d1})$-$(\ref{d4})$ and $(\ref{torsion})$, we
can change the form of $\bar{R}_{A'}$ and find,  \fa
\bar{R}_{A}&=&(\textstyle{\frac{-i}{g^2l}}\epsilon^{abc}f_{bc}^{A'}
+\textstyle{\frac{G\sqrt{\Lambda}}{3l}}\pi^{aA'})e_{aA'A}=D_a\pi^a_A
+\textstyle{\frac{G\sqrt{\Lambda}}{3}}\pi^a_{AB}\psi^B_a\nonumber\\&=&L_A
-\frac{2G\sqrt{\Lambda}}{3}\pi^a_{AB}\psi^B_a,\nonumber\\
\bar{L}_{A'}&=&\bar{D}_{a}\pi_{A'}^{a}+\textstyle{G\sqrt{\Lambda}}
\pi_{A'B'}^{a}\psi_{a}^{B'}
=(\textstyle{\frac{i}{g^2l}}\epsilon^{abc}f_{bc}^{A}+\textstyle{G\sqrt{\Lambda}}
\pi^{aA})e_{aAA'}\nonumber\\&=&R_{A'}
+\frac{2G\sqrt{\Lambda}}{3l}\pi^{aA}e_{aAA'}.\label{GA} \ffa To
show this, we only need consider the key term and find it can be
changed into the following expression, \fa
\textstyle{\frac{-i}{g^2l}}\epsilon^{abc}f_{bc}^{A'}e_{aA'A}
&=&\textstyle{\frac{-i}{g^2l}}\epsilon^{abc}[d_b\chi_c^{A'}+A_b^{A'B'}\chi_{cB'}]e_{aA'A}\nonumber\\
&=&\textstyle{\frac{-i}{g^2l}}\epsilon^{abc}[d_b\chi_c^{A'}e_{aA'A}]+
\textstyle{\frac{-i}{g^2l}}\epsilon^{abc}\chi_c^{A'}(d_be_{aA'A}+A_{bA'}^{B'}e_{aB'A})\nonumber\\&=&
D_b\pi_A^b-A_{bA}^B\pi_B^b+\textstyle{\frac{-i}{g^2l}}\epsilon^{abc}\chi_c^{A'}\bar{D}_be_{aA'A}\nonumber\\
&=&D_b\pi_A^b-A_{bA}^B\pi_B^b+\textstyle{\frac{-i}{g^2l}}\epsilon^{abc}\chi_c^{A'}[-A_{bA}^Be_{aA'B}-
\frac{G\sqrt{\Lambda}}{2l}\chi_{bA'}\psi_{aA}]\nonumber\\&=&D_b\pi_A^b-
\frac{i\sqrt{\Lambda}}{2}\epsilon^{abc}\chi_c^{A'}\chi_{bA'}\psi_{aA}\nonumber\\
&=&D_b\pi_A^b.\ffa Combining these equations together we  find:
\fa G^L_{A}
&=&2(D_{a}\pi_{A}^{a}+\frac{\scriptstyle{2G\sqrt{\Lambda}}}{3}\pi_{AB}^{a}\psi_{a}^{B})=0,
\label{super3}\\ G^R_{A'}&=&2(\textstyle{\frac{i}{g^2l}}
\epsilon^{abc}f_{bc}^{A}+\textstyle{\frac{2G\sqrt{\Lambda}}{3l}}
\pi^{aA})e_{aAA'}=0.\label{super4} \ffa

They are conjugate of each other indeed! We may also note that if
we write \f G^R_{A'}={ N_{A'}^{A} \over \sqrt{h}}G^R_{A}, \ff we
have \f G^R_{A}=2(\textstyle{\frac{i}{g^2l}}
f_{bc}^{B}+\textstyle{\frac{2G\sqrt{\Lambda}}{3l}}
\pi^{aB}\epsilon_{abc})\pi^{b}_{BD}\pi^{cDA}. \ff

If we consider the Poisson brackets between the $G_{AB}$ and
$G^L_A$, we find that they does form a close algebra for
$Osp(1|2)$ as in the case of chiral supergravity \f \{G(\lambda),
G(\lambda')\}=G([\lambda, \lambda']),\ff \f \{G^L(\eta),
G(\lambda)\}=G^L([\eta,\lambda]),\ff \f \{G^L(\eta),
G^L(\eta')\}=G([\eta, \eta']),\ff where $G$ and $G^L$ are the
constrained functional smeared on the three dimensional space. The
same is true for the combination of the $G_{A'B'}$ and $G_{A'}$.
So all these equations form the super Gauss's law which generates
the $Osp(1|2)_{L}\oplus Osp(1|2)_R$ gauge transformations of the
canonical variables.\\ In this sense we can say our model is
combining two copies of the chiral supergravities into one.

We still have to discuss the commutator of $G^{A}$ with $G^{A'}$.
This requires more work as it involves the hamiltonian and
diffeomorphism constraints.

\subsection{The hamiltonian and diffeomorphism
Constraints}

We come finally to the constraint associated with the Lagrangian
multiplier $e_{0AA'}$. As in \cite{hologr} these will contain the
Hamiltonian and diffeomorphism constraints. We have \fa
G^{AA'}:&=&\textstyle{\frac{2}{l^2}}\pi^{aAB}e_{aB}^{A'}-
\textstyle{\frac{1}{l^2}}\pi^{aA}\chi_{a}^{A'}
-\textstyle{\frac{2i}{l^2}}\epsilon^{abc}\lambda_{ab}^{AB}e_{cB}^{A'}-
\textstyle{\frac{i}{l}}\epsilon^{abc}
\lambda_{ab}^{A}\chi_{c}^{A'}\nonumber\\ & &
+\textstyle{\frac{2}{l^2}}\pi^{aA'B'}e_{aB'}^{A}-
\textstyle{\frac{1}{l^2}}\pi^{aA'}\psi_{a}^{A}
+\textstyle{\frac{2i}{l^2}}\epsilon^{abc}\lambda_{ab}^{A'B'}e_{cB'}^{A}+
\textstyle{\frac{i}{l}}\epsilon^{abc}
\lambda_{ab}^{A'}\psi_{c}^{A}\nonumber\\&=&0.\label{H} \ffa To
understand this constraint better, we try to change it into a new
form in which $G_{AA'}$ is divided into two conjugate parts
$(C_{AB},C_{A'B'})$.

\f G^{AA'}={2 \over G} \left ( C^{AB}{ N_B^{A'}\over \sqrt{h}}+
C^{A'B'}{N_{B'}^A \over \sqrt{h}}\right ). \ff To find the forms
of $C^{AB}$ and $C^{A'B'}$ we need only to plug all the solutions
$(\ref{d1})$-$(\ref{d4})$ and $(\ref{l1})$-$(\ref{l4})$ into
($\ref{H}$), then we find $C^{AB}$ and $C^{A'B'}$ have the
following form \fa
C^{AB}&=&\textstyle{\frac{G^2\Lambda}{3}}\epsilon_{abc}
\pi^{aAC}\pi^{bBD}\pi_{CD}^{c}-2i\pi^{aBC}\pi_{DC}^{b}f_{ab}^{AD}
-i\textstyle{\frac{G\sqrt{\Lambda}}{l}}\pi^{aBC}\pi_{DC}^{b}\psi_a^A\psi_b^D\nonumber\\
&&+\textstyle{\frac{G\sqrt{\Lambda}}{3}g^2}\epsilon_{abc}
\pi^{aA}\pi^{bBC}\pi_{C}^{c}-2i\pi^{aBC}\pi_{C}^{b}f_{ab}^{A},
\ffa \fa C^{A'B'}&=&\textstyle{\frac{G^2\Lambda}{3}}\epsilon_{abc}
\pi^{aA'C'}\pi^{bB'D'}\pi_{C'D'}^{c}+2i\pi^{aB'C'}\pi_{D'C'}^{b}f_{ab}^{A'D'}
+i\textstyle{\frac{G\sqrt{\Lambda}}{l}}\pi^{aB'C'}\pi_{D'C'}^{b}\chi_a^{A'}\chi_b^{D'}\nonumber\\
&&+\textstyle{\frac{G\sqrt{\Lambda}}{3}g^2}\epsilon_{abc}
\pi^{aA'}\pi^{bB'C'}\pi_{C'}^{c}+2i\pi^{aB'C'}\pi_{C'}^{b}f_{ab}^{A'}.
\ffa

As is the case in general relativity \cite{hologr}, it turns out
that the four constraints $C^{AB}$ contain both the diffeomorphism
and hamiltonian constraints of the supergravity. If we only
consider the terms only involving the variables $\pi^a_{AB}$ and
$F_{abAB}$, then the constraints go back to those of the quantum
general relativity \cite{hologr}.

\section{The boundary theory}

We are now ready to focus on the construction of the boundary
theory. We follow the formulation given in \cite{hologr} for
general relativity with Lorentzian signature.  This is related to
the Euclidean formulation given in \cite{linking}, but differs
from that in several significant aspects.

We study first the addition of a boundary to the action and
equations of motion. Once this is understood we will proceed to
their expression in the Hamiltonian formulation.

\subsection{Boundary terms in the action and equations
of motion}

In the presence of the boundary, the ordinary variations of the
action are not well defined unless boundary terms  are added to
the action and some boundary conditions are imposed. These must be
chosen so as to cancel the boundary term in the variation of the
action, so that the total action is functionally differentiable.
We will assume that $\cal M$ has the topology of $R \times
\Sigma$, so that $\partial {\cal M} \approx R \times
\partial
\Sigma$, where $\partial \Sigma$ is a compact two manifold.  We
will assume here that $\partial \Sigma \approx S^2$, but it is not
difficult to consider the more general case.

It's very interesting that the total action (\ref{action0}) can be
supplemented by the $Osp(1|2)_{L} \oplus Osp(1|2)_{R}$ super
Chern-Simons term with the appropriate boundary condition \fa
S_{cs}&=&\textstyle{\frac{ik}{4\pi}}\int_{\partial
M}Y_{cs}(A_{AB},\psi_A)-\textstyle{\frac{ik}{4\pi}}\int_{\partial
M}Y_{cs}(A_{A'B'},\chi_{A'})\nonumber\\&=&
\textstyle{\frac{ik}{4\pi}}\int_{\partial M}(A^{AB}\wedge
\tilde{F}_{AB}+\textstyle{\frac{2}{3}}A\wedge A\wedge
A+\textstyle{\frac{G\sqrt{\Lambda}}{2l}}\psi^A\wedge\tilde
{F}_A)\nonumber\\&&-\textstyle{\frac{ik}{4\pi}}\int_{\partial
M}(A^{A'B'}\wedge \tilde{F}_{A'B'}+\textstyle{\frac{2}{3}}A'\wedge
A'\wedge
A'+\textstyle{\frac{G\sqrt{\Lambda}}{2l}}\chi^{A'}\wedge\tilde{F}_{A'}),
\label{bterm} \ffa

Where $(\tilde {F}_{AB},\tilde{F}_A)$ and $(\tilde
{F}_{A'B'},\tilde{F}_{A'})$ are components of the curvature of the
one form connection of $Osp(1|2)_{L}$ and $Osp(1|2)_{R}$
respectively. They are of the form \fa \tilde{F}_{AB}&=&f_{AB}+
\textstyle{\frac{G\sqrt{\Lambda}}{2l}}\psi_A\wedge\psi_B,
\nonumber\\ \tilde{F}_{A}&=& D\psi_{A}= f_A,\nonumber\\
\tilde{F}_{A'B'}&=&f_{A'B'}+
\textstyle{\frac{G\sqrt{\Lambda}}{2l}}\chi_{A'}\wedge\chi_{B'},
\nonumber\\ \tilde{F}_{A'}&=& \bar{D}\chi_{A'}= f_{A'}.\ffa If we
take the variation of the total action \fa \delta S_{tot}&=&\delta
S_{BF}+\delta S_{CS}\nonumber\\ &=&\int_{M}(...)+\int_{\partial
M}\delta A^{AB}\wedge(\frac{1}{g^2}B_{AB}-\frac{k}{2\pi}
\tilde{F}_{AB})+\delta\psi^A\wedge[\frac{1}{g^2}
B_A-\frac{k}{2\pi}\textstyle{\frac{G\sqrt{\Lambda}}{2}}(
\tilde{F}_A+A_{AB}\wedge\psi^B)]\nonumber\\&-&\int_{\partial
M}\delta A^{A'B'}\wedge(\frac{1}{g^2}B_{A'B'}-\frac{k}{2\pi}
\tilde{F}_{A'B'})+\delta\chi^{A'}\wedge[\frac{1}{g^2}B_{A'}-\frac{k}{2\pi}
\textstyle{\frac{G\sqrt{\Lambda}}{2}}(\tilde{F}_{A'}+A_{A'B'}\wedge\chi^{B'})],
\ffa we require that the boundary terms vanish.  In order to
induce a local boundary theory, we require that the integrand of
the boundary term in the variation vanish.  This leads to the
following conditions, \f \epsilon^{abc}\delta
A^{AB}_c(\frac{1}{g^2} B_{abAB}-\frac{k}{2\pi}\tilde{F}_{abAB})=0,
\ff \f \epsilon^{abc}\delta \psi^{A}_c[\frac{1}{g^2}
B_{abA}-\frac{k}{2\pi}\textstyle{\frac{G\sqrt{\Lambda}}{2}}(\tilde{F}_{abA}+A_{aAB}\wedge\psi_b^B)]=0,
\label{SB1}\ff \f \epsilon^{abc}\delta A^{A'B'}_c(\frac{1}{g^2}
B_{abA'B'}-\frac{k}{2\pi}\tilde{F}_{abA'B'})=0, \ff \f
\epsilon^{abc}\delta \chi^{A'}_a[\frac{1}{g^2}
B_{abA'}-\frac{k}{2\pi}\textstyle{\frac{G\sqrt{\Lambda}}{2}}(\tilde{F}_{abA'}+A_{aA'B'}\wedge\chi_b^{B'})]=0.
\ff There are several ways to satisfy these conditions. Taking the
boundary term involving $\delta A_{AB}$ as an example, we find

\begin{itemize}
\item
One can fix the variable $\cal A\mit_a$ on the boundary: \f \delta
A_a^{AB}|_{\partial {\cal M}}=0. \label{w1} \ff

\item One can take the self-dual conditions for the
curvature $ \tilde{F}_{abAB}$: \f \left ( \tilde{F}_{abAB}-
\frac{2\pi}{g^2k} B_{abAB} \right )_{\partial {\cal M}} =0.
\label{w2} \ff
\item We can consider some combination of these two
conditions in which (\ref{w1}) is imposed on some components of
the connection while (\ref{w2}) is imposed on the others.

\end{itemize}

The first case is simple, but leads to a reduction of the local
gauge invariance on the boundary. The next simplest case would be
the second (\ref{w2}).  This is what was done for the Euclidean
theory in \cite{linking}. An advantage of this condition is that
it leads to no reduction of the local gauge symmetry on the
boundary. However, in the Lorentzian case, (\ref{w2}) is too
strong, as it can be shown that all of its solutions correspond to
a surface wiggling in a  fixed $AdS$ space time.  By counting
there is one degree of freedom, but this turns out to correspond
to local longitudinal motions of the surface.

This problem does not arise in the Euclidean case studied in
\cite{linking} because the left and right handed components of the
connection are independent.  It is a consequence of the fact that
in the Lorentzian theory they are instead related by the reality
conditions, which require that they are complex conjugates of each
other.

If our boundary conditions are to allow an infinite dimensional
space of solutions we must then loosen the self-dual boundary
conditions. We do this as follows.

We begin by imposing a time slicing of the boundary, which
corresponds to some choice of time coordinate $t$ on $\partial
{\cal M}$. We then restrict attention to a spatial slice of the
boundary at fixed $t=$constant. We proceed by specifying a set of
preferred points on the spatial boundary, which will be called
punctures.  They will correspond to points where the spin network
states meet the boundary in the quantum theory.  Each puncture is
surrounded by a region of the spatial boundary. In each region we
introduce local coordinates $(r,\theta)$ which are angular
coordinates with the puncture at the origin.  These can then be
joined yielding a single coordinate patch on the whole punctured
sphere, which reduces to an angular coordinate system in the
neighborhood of each puncture. Bringing back $t$ we then have a
coordinate system $(r,\theta, t)$ on the whole of ${\partial M}$
minus the world lines of the punctures.

Then we put the following conditions for the components of the
fields

\f \frac{1}{g^2}B_{r\theta AB}-\frac{k}{2\pi}\tilde{F}_{r\theta
AB}=0, \label{b1} \ff \f \frac{1}{g^2}B_{\theta
tAB}-\frac{k}{2\pi}\tilde{F}_{\theta tAB}=0, \label{b2} \ff \f
\delta A_{\theta}=0. \label{b3} \ff The first two equations,
(\ref{b1}) and (\ref{b2}) correspond to  imposing the self-dual
boundary conditions on the components $F_{r\theta}$ and
$F_{t\theta}$. But instead of constraining the $F_{rt}$ component
we impose the third condition (\ref{b3}).  The radial component of
the connection, $\cal A\mit_r$, is thus unrestricted and is free
to evolve with time.

It can be checked that when these conditions have been imposed the
bulk theory is not trivial and has an infinite number of
solutions.

We can then write the boundary conditions in terms of the
component fields, yielding, \f f_{r\theta
AB}=\frac{1}{l^2}\frac{2\pi}{g^2k}e_{rAA'}\wedge e_{\theta
B}^{A'}-\frac{G\sqrt{\Lambda}}{2l}\psi_{rA}\wedge\psi_{\theta
B},\ff \f f_{r\theta
A}=\frac{1}{l}\frac{2\pi}{g^2k}e_{rAA'}\wedge\chi_{\theta}^{A'},
\ff \f f_{t\theta AB}=\frac{1}{l^2}\frac{2\pi}{g^2k}e_{tAA'}\wedge
e_{\theta
B}^{A'}-\frac{G\sqrt{\Lambda}}{2l}\psi_{tA}\wedge\psi_{\theta
B},\ff \f f_{t\theta
A}=\frac{1}{l}\frac{2\pi}{g^2k}e_{tAA'}\wedge\chi_{\theta}^{A'},
\ff \f \delta A_{\theta AB}=0\;\;\;\;\;\;\;\delta \psi_{\theta
A}=0. \ff Consistency with the field equations leads to
relationship between the constant $k$ and the cosmological
constant, \f G^2\Lambda=\frac{6\pi}{k}. \ff

\subsection{Boundary terms in the canonical theory}

Next we describe the expression of the boundary terms and
conditions on the canonical theory.   The first effect to take
into account is that the boundary terms alter the definition of
the momenta of the fields, due to the fact that (\ref{bterm})
contains time derivatives, \f
\pi_{AB}^{a}:=\textstyle{\frac{-i}{g^2}}\epsilon^{abc}B_{bcAB}(x)+
\textstyle{\frac{ik}{4\pi}}\int
d^{2}S(\sigma)\epsilon^{ab}A_b^{AB}\delta^3(x, S (s)),\ff \f
\pi_{A'B'}^{a}:=\textstyle{\frac{i}{g^2}}\epsilon^{abc}B_{bcA'B'}(x)+
\textstyle{\frac{-ik}{4\pi}}\int
d^{2}S(\sigma)\epsilon^{ab}A_b^{A'B'}\delta^3(x,S (s)),\ff \f
\pi_{A}^{a}:=\textstyle{\frac{-i}{g^2}}\epsilon^{abc}B_{bcA}(x)+
\textstyle{\frac{ik}{4\pi}}\int
d^{2}S(\sigma)\epsilon^{ab}\textstyle{\frac{G\sqrt{\Lambda}}{2l}}
\psi_b^{A}\delta^3(x, S (s)),\ff \f
\pi_{A'}^{a}:=\textstyle{\frac{i}{g^2}}\epsilon^{abc}B_{bcA'}(x)+
\textstyle{\frac{-ik}{4\pi}}\int
d^{2}S(\sigma)\epsilon^{ab}\textstyle{\frac{G\sqrt{\Lambda}}{2l}}
\chi_b^{A'}\delta^3(x, S (s)),\ff where $\epsilon^{ab}$ is the
area element induced by the volume element on the surface with \f
\epsilon^{ab}:=\epsilon^{abc}n_c.\ff The Gaussian constraints can
be expressed as: \f G^{AB}=D_ap^{aAB}+q^{AB},\ff Where \f
p^{aAB}:=\textstyle{\frac{-i}{g^2}}\epsilon^{abc}B_{bc}^{AB}(x)+
\textstyle{\frac{ik}{2\pi}}\int
d^{2}S(\sigma)\epsilon^{ab}A_b^{AB}\delta^3(x, S (s)),\ff and \f
q^{AB}:=\textstyle{\frac{i}{g^2l}}\epsilon^{abc}
B_{ab}^{A}\psi_c^B+\textstyle{\frac{ik}{2\pi}}\int
d^{2}S(\sigma)\epsilon^{ab}\textstyle{\frac{G\sqrt{\Lambda}}{2l}}
\psi_a^{A}\psi_b^{B}\delta^{3}(x,\delta(s)).\ff The Gauss
functional can be obtained by smearing the constraint on the
spatial manifold with the appropriate boundary terms, \fa
G(\Lambda_{AB})&=&\int_{\Sigma}\Lambda_{AB}G^{AB}=\int_{\Sigma}\Lambda_{AB}[
D_{a}(\textstyle{\frac{-i}{g^2}}
\epsilon^{abc}B_{bc}^{AB})-\textstyle{\frac{-i}{g^2l}}\epsilon^{abc}
B_{ab}^{A}\psi_c^B]\nonumber\\ &&-\frac{ik}{2\pi}
\int_{\Sigma}{\Lambda_{AB}}\int_{}
{d^{2}S(\sigma)\epsilon^{ab}D_{a}A_b^{AB}\delta^3(x,S(s))}\nonumber\\
&&+\frac{ik}{2\pi}\int_{\Sigma}\Lambda_{AB}\int
d^{2}S(\sigma)\epsilon^{ab}\textstyle{\frac{G\sqrt{\Lambda}}{2l}}
\psi_a^{A}\psi_b^{B}\delta^{3}(x,S(s))\nonumber\\
&=&-\int_{\Sigma}{\textstyle{\frac{i}{g^2}}
\epsilon^{abc}B_{bc}^{AB}D_{a}\Lambda_{AB}}
-\int_{\Sigma}\Lambda_{AB}\textstyle{\frac{-i}{g^2l}}\epsilon^{abc}
B_{ab}^{A}\psi_a^{B}\nonumber\\
&&+\int_{\partial\Sigma}ds\Lambda_{AB} [\textstyle{\frac{-i}{g^2}}
\epsilon^{abc}B_{bc}^{AB}n_a+\textstyle{\frac{ik}{2\pi}}
\epsilon^{ab}f_{ab}^{AB}+\textstyle{\frac{ik}{2\pi}}
\textstyle{\frac{G\sqrt{\Lambda}}{2l}}\epsilon^{ab}\psi_a^{A}\psi_b^{B}]
\ffa In the canonical theory the constraint then has a
 boundary
term: \f G(\Lambda_{AB})|_b=\int_{\partial\Sigma}ds\Lambda_{AB}
[\textstyle{\frac{-i}{g^2}}
\epsilon^{abc}B_{bc}^{AB}n_c+\textstyle{\frac{ik}{2\pi}}\epsilon^{ab}(f_{ab}^{AB}+
\textstyle{\frac{G\sqrt{\Lambda}}{2l}}\psi_a^{A}\psi_b^{B})]. \ff
This leads to the canonical form of the boundary terms. \f
\frac{1}{g^2} B_{abAB}=\frac{k}{2\pi}\tilde{F}_{abAB}. \ff

Notice that this corresponds with one of the components of the
boundary terms we imposed in the lagrangian theory (\ref{b1}). The
boundary conditions (\ref{b2}) do not arise in the same way, as
they involve the time components of fields, which become lagrange
multipliers in the canonical theory.  The last components of the
lagrangian theory (\ref{b3}) persist in the Hamiltonian theory.
These in fact correspond to the fact that in the quantum theory
the labels of the punctures are fixed and do not evolve in time.

In a similar way, the boundary term on the supersymmetric
constraints can be derived as follows: \f G^{A}=D_ap^{aA}+q^{A}\ff
where\f
p^{aA}:=\textstyle{\frac{-i}{g^2}}\epsilon^{abc}B_{bc}^A(x)+
\textstyle{\frac{ik}{2\pi}}\int
d^{2}S(\sigma)\epsilon^{ab}\textstyle{\frac{G\sqrt{\Lambda}}{2l}}
\psi_b^{A}\delta^3(x, S (s)),\ff \f
q^A:=\textstyle{\frac{i}{g^2l}}\textstyle{\frac{2G\sqrt{\Lambda}}{3l}}\epsilon^{abc}
B_{ab}^{AB}\psi_{cB}+\textstyle{\frac{ik}{2\pi}}\int
d^{2}S(\sigma)\epsilon^{ab}\textstyle{\frac{G\sqrt{\Lambda}}{2l}}
A_a^{AB}\psi_{bB}\delta^{3}(x,S(s)),\ff and the supersymmetric
functional is  \fa
G(\Lambda_{A})&=&\int_{\Sigma}\Lambda_{A}G^{A}=\int_{\Sigma}{\Lambda_{A}[
D_{a}(\textstyle{\frac{-i}{g^2}}
\epsilon^{abc}B_{bc}^{A})-\textstyle{\frac{-i}{g^2l}}\textstyle{\frac{2G\sqrt{\Lambda}}{3l}}\epsilon^{abc}
B_{ab}^{AB}\psi_{cB}]}\nonumber\\
&&-\frac{ik}{2\pi}\frac{G\sqrt{\Lambda}}{2l}
\int_{\Sigma}{\Lambda_{A}}\int
d^{2}S(\sigma)\epsilon^{ab}D_{a}\psi_b^{A}\delta^3(x,S(s))\nonumber\\
&&+\frac{ik}{2\pi}\int_{\Sigma}\Lambda_{A}\int
d^{2}S(\sigma)\epsilon^{ab}\textstyle{\frac{G\sqrt{\Lambda}}{2l}}
A_b^{AB}\psi_{aB}\delta^{3}(x,S(s))\nonumber\\
&=&-\int_{\Sigma}{\textstyle{\frac{-i}{g^2}}
\epsilon^{abc}B_{bc}^{A}D_{a}\Lambda_{A}}
-\int_{\Sigma}\Lambda_{A}\textstyle{\frac{-i}{g^2l}}\textstyle{\frac{2G\sqrt{\Lambda}}{3l}}\epsilon^{abc}
B_{ab}^{AB}\psi_{cB}\nonumber\\
&&+\int_{\partial\Sigma}ds\Lambda_{A} [\textstyle{\frac{-i}{g^2}}
\epsilon^{abc}B_{bc}^{A}n_a-\textstyle{\frac{ik}{2\pi}}
\textstyle{\frac{G\sqrt{\Lambda}}{2l}}\epsilon^{ab}(f_{ab}^{A}+
A_a^{AB}\psi_{bB})]. \label{SB2}\ffa Comparing $(\ref{SB2})$ with
$(\ref{SB1})$, we have \f\frac{1}{g^2}
B_{abA}=\frac{k}{2\pi}\textstyle{\frac{G\sqrt{\Lambda}}{2}}(\tilde{F}_{abA}+A_{aAB}\wedge\psi_b^B).
\ff From above consideration, we find both Gauss constraint and
the supersymmetric constraint are well defined on the spatial
surface if we impose the self-dual conditions for the curvatures.

\section{Quantization of the theory}

We are finally ready to discuss the quantization of supergravity,
in the presence of boundary conditions (\ref{b1}-\ref{b3}). We
follow the methods of loop quantum gravity
\cite{sn1,sn2,spain,lp1,lp2}, extended to theories with finite
boundaries in \cite{linking,hologr}. Two important parts of the
theory are the spin network basis \cite{sn1,sn2} which we extended
to supergravity in \cite{superspin}, and the Chern-Simons state
\cite{kodama,chopinlee}, which was extended to supergravity in
\cite{GSU1}. The combination yields framed or quantum-deformed
supersymmetric spin networks, which are constructed from the
representation theory of quantum deformed $Osp(1|2)$ \cite{q-osp}.

The method of treating the relationship between the boundary and
bulk theory we use was developed in \cite{linking} and then
extended to theories with Lorentzian signature in \cite{hologr}.
Here we describe only the results, the reader is referred to
\cite{linking} and \cite{hologr} for detail.

There is  a well known non-perturbative procedure to deal with the
quantization of the bulk theory in the case of general relativity
 \cite{sn1, sn2, spain, lp1, lp2}. A set of complete
but independent basis of the quantum states can be constructed  by
means of spin networks. It is a trivalent graph $\Gamma$ in which
each node join some links. Associated every link, there is a spin
$j_i$ which satisfies some conditions at the node, which is also
labeled by an intertwiner $\nu_e$. Recently we extended these
techniques to the N=1 chiral supergravity with super Lie algebra
$Osp(1|2)$
 \cite{superspin}. It turns out that an analogous
supersymmetric spin networks can be established for the theory. In
particular in the basis of super spin networks the spectrum of the
area operator can be computed out and partly diagonalized. For
tri-valent spin networks its eigenvalues have a discrete form
given by \cite{superspin}:

\f \hat{A}|\Gamma^{sg},n_i,v_e\rangle=\sum_i l_{p}^2
\sqrt{j_i(j_i+\frac{1}{2})}| \Gamma^{sg},n_i,v_e\rangle.
\label{area}\ff Where $l_{p}$ is the Planck length and
$j_i={n_i\over 2}$. We need modify the construction slightly to
fit the constrained super $BF$ theory, which double covers over
the chiral supergravity. The configuration space of the theory is
the modular space of $Osp(1|2)\oplus Osp(1|2)$. The basis for the
Hilbert space can be constructed in terms of the super spin
networks for the super algebra $Osp(1|2)_L\oplus Osp(1|2)_R$. Then
the links of the networks are labeled by conjugate pairs of
spins$(j_L,j_R)$ and the nodes are labeled by pairs of
intertwiners$(\nu_L,\nu_R)$. There is a similar case as in quantum
general relativity that we need impose the balanced conditions to
the pairs of$(j_L,j_R)$ and$(\nu_L,\nu_R)$ to coincide with the
reality condition of the theory. These conditions are $j_L=j_R$
and $\nu_L=\nu_R$.

All the construction above can be extended without difficulty to
the quantum deformed version of the superalgebra, $U_q(Osp(1|2))$,
whose representation are finite dimensional and bounded by the
generic $q$. \footnote{The quantum superalgebra and supergroups
$U_q(osp(1|2))$ are investigated in several places $\cite{q-osp}$.
and its structure of finite dimensional representation are well
established.} The q-deformed formalism of the supergroup give some
modification to the spectrum of the area but having a similar
expression as the ordinary one, \f
\hat{A}|\Gamma^{sg},n_i,v_e\rangle=\sum_i l_{p}^2
\sqrt{[j_i][j_i+\frac{1}{2}]}| \Gamma^{sg},n_i,v_e\rangle, \ff
where \f [j]:=\frac{q^j-q^{-j}}{q-q^{-1}}.\ff

If we impose all the constraints on the state space quantum
mechanically and require the results of their action vanish, we
could find a set of physical states which actually is the super
Chern-Simons states,
 \f
\Psi (\Gamma)=\int D\cal A\mit e^{
\textstyle{\frac{ik}{4\pi}}(Y_{cs}(A_{AB},\psi_A)-Y_{cs}(A_{A'B'},\chi_{A'}))}.
\ff

It allows us to testify the Bekenstein bound in a quantum
mechanical sense. We consider the spatial n-punctured boundary
with finite area. Associated to every puncture , there is a
half-integer $j_{\alpha}$ which takes values from $1/2$ to $k/2$,
where k is the coupling constant of the Chern-Simons theory. Then
the total Hilbert space is constructed by the direct sum of the
state spaces of all $Osp(1|2)\oplus Osp(1|2)$ Chern-Simons
theories. Here we pay more attention to the contribution of the
boundaries. As we know, for every set of the punctures the Hilbert
space of the topological field theory has finite number of degrees
of freedom. In our case it is contributed by the conformal blocks
for the n punctures of the $Osp(1|2)\bigoplus Osp(1|2)$ WZW model.
We find: \f \cal {H}\mit_{j_1,...,j_n}=\cal
{V}\mit_{j_1,...,j_n}^{L}\bigoplus \cal
{V}\mit_{j_1,...,j_n}^{R},\ff and the dimensions of the space is:
\f dim(\cal {V}\mit_{j_1,...,j_n}^{L}\bigoplus \cal
{V}\mit_{j_1,...,j_n}^{R})=\prod_i(4j_i+1).\label{dim}\ff

Then based on this result and the holography hypothesis, we can
construct the total physical state space as the summation of all
the possible states of the topological field theories: \f \cal
{H}\mit^{phys}=\sum_n\sum_{j_1,...,j_n}\cal
{H}\mit_{j_1,...,j_n}\ff Using $(\ref{area})$ and $(\ref{dim})$,
and choosing the constant c, we find the Bekenstein bound is
satisfied: \f dim(\cal {V}\mit_{j_1,...,j_n}^{L}\bigoplus \cal
{V}\mit_{j_1,...,j_n}^{R})\le
exp\{{\textstyle{\frac{cA[j_1,...,j_n]}{G\hbar}}}\}.\ff

\section{Conclusions}

In this paper we have extended to the case of $N=1$ supergravity
the basic facts about finite area boundary conditions worked out
previously for quantum general relativity in
\cite{linking,hologr}. Here we have treated the case of timelike
boundaries following \cite{hologr}, it is straightforward also to
find the same results for the Euclidean theory with cosmological
constant by restricting to the left handed sector and then
Euclideanizing, as in \cite{linking}.  A natural extension of the
Chern-Simons boundary conditions studied in \cite{linking} to null
boundaries, suitable for the descriptions of horizons, has been
worked out in \cite{isolated}, following observations made in
\cite{kirill}. Following the results given here it should be
straightforward also to extend these results to supergravity.
Another important thing yet to be done is the construction of
quasi-local operators on the boundary which represent the
hamiltonian and supersymmetry charges.  It may be hoped that this
may resolve issues concerning the positivity of the Hamiltonian in
loop quantum gravity\cite{positive}.

The next works in this series will extend these results to $N=2$
supergravity, which should make possible the detailed description
of $BPS$ states and $BPS$ black holes in the langauge of loop
quantum gravity.  This should make possible the direct comparison
of the results on black hole state counting coming from string
theory and loop quantum gravity. Another application of these
results may be to an explicit construction of the boundary
conformal field theory in the AdS/CFT conjecture in $3+1$
dimensions.

As a final remark we emphasize the natural way in which the
Bekenstein bound is satisfied in quantum general relativity given
only the imposition of the appropriate Chern-Simons boundary
conditions\cite{linking}. We see here that this extends naturally
to supergravity.   This appears to make possible formulations of
the holographic principle at the background independent
level\cite{louis-holo,weakholo,weakstong,mpaper}.

\section*{ACKNOWLEDGEMENTS}
We are grateful to Abhay Asthtekar, Roger Penrose, Michael
Reisenberger and Carlo Rovelli for comments on various aspects of
this work. Also both of us would like to thank the theoretical
physics group at Imperial college for their hospitality during
this last year. This work is supported by the NSF through grant
PHY95-14240 and a gift from the Jesse Phillips Foundation.

\end{document}